\documentclass[reprint,amsmath,amssymb,aps,prb,showpacs,floatfix,byrevtex]{revtex4-1}

\usepackage{amsfonts}
\usepackage{bm}
\usepackage{dcolumn}
\usepackage{graphicx}
\usepackage{wasysym}

\begin{document}

\preprint{Draft --- not for distribution}

%
%
\title{Electronic correlations and unusual superconducting response in the
  optical properties of the iron-chalcogenide
  \boldmath FeTe$_{0.55}$Se$_{0.45}$ \unboldmath}
\author{C. C. Homes}
\email{homes@bnl.gov}
\author{A. Akrap}
\author{J. S. Wen}
\author{Z. J. Xu}
\author{Z. W. Lin}
\author{Q. Li}
\author{G. D. Gu}
\affiliation{Condensed Matter Physics and Materials Science Department,
  Brookhaven National Laboratory, Upton, New York 11973, USA}%
\date{\today}

%
%
\begin{abstract}
The in-plane complex optical properties of the iron-chalcogenide superconductor
FeTe$_{0.55}$Se$_{0.45}$ have been determined above and below the critical
temperature $T_c = 14$~K.  At room temperature the conductivity is described
by a weakly-interacting Fermi liquid; however, below 100~K the scattering rate
develops a frequency dependence in the terahertz region, signalling the
increasingly correlated nature of this material.
We estimate the dc conductivity $\sigma_{dc}(T\gtrsim T_c) \simeq 3500\pm 400$
$\Omega^{-1}$cm$^{-1}$ and the superfluid density $\rho_{s0} \simeq 9\pm 1\times
10^6$~cm$^{-2}$, which places this material close to the scaling line $\rho_{s0}/8
\simeq 8.1\,\sigma_{dc}T_c$ for a BCS dirty-limit superconductor.  Below $T_c$ the
optical conductivity reveals two gap features at $\Delta_{1,2} \simeq 2.5$ and
5.1~meV.
\end{abstract}
%
%
%
%
%
%
%
%
\maketitle

%
%
%
The surprising discovery of superconductivity in the iron-arsenic LaFeAsO$_{1-x}$F$_x$
(``1111'') pnictide compound has prompted an intense investigation of this class of
materials.\cite{kamihara08,ishida09}  The critical temperature $T_c$ may be increased
above 50~K through rare-earth substitutions.\cite{ren08}  While the mechanism for
superconductivity in many metals and alloys is mediated by lattice vibrations,\cite{bcs}
the high values for $T_c$ and the strong interplay between the magnetism and the
lattice suggest that the superconductivity in this class of materials is not phonon
mediated.\cite{boeri08}
In addition to searching for higher values of $T_c$ in the 1111 family of
materials, considerable effort has been made looking for superconductivity in
other structurally-simpler Fe-based systems.  In metallic BaFe$_2$As$_2$
the application of pressure yields $T_c \simeq 29$~K, while Co- and Ni-doping
yields $T_c \simeq 23$~K at ambient pressure.\cite{alireza09,sefat08,li09}
Superconductivity has also been observed in the As-free iron-chalcogenide FeSe
compound with $T_c = 8$~K, which increases to $T_c = 27$~K with the application
of pressure.\cite{hsu08,mizuguchi08}  By introducing Te, the critical temperature in
FeTe$_{1-x}$Se$_{x}$ at ambient pressure reaches a maximum $T_c = 14$~K for $x=0.45$.
Despite these structural differences, the band structure of these materials is
similar, with a minimal description consisting of an electron band $(\beta)$ at
the M point and a hole band ($\alpha)$ at the center of the Brillouin zone.\cite{raghu08}
There have been a number of studies of the Fe$_{1+x}$Te and FeTe$_{1-x}$Se$_{x}$ materials,
including transport,\cite{fang08,chengf09,sales09,taen09} tunneling\cite{kato09} and
angle-resolved photoemission,\cite{xiay09,nakayama09,tamai10,leesh09} with particular
emphasis placed on the magnetic properties.\cite{leesh09,wen09,qui09,khasanov09,han09,bao09}
While the optical properties of the superconducting iron-pnictide materials have been
investigated in some detail,\cite{hu09,yang09,kim09,nakajima10,heumen09,gorshunov10,wu10}
in comparison the iron-chalcogenide materials remain relatively unexplored.\cite{chengf09}

%
%
In this work we examine the in-plane complex optical properties of superconducting
FeTe$_{0.55}$Se$_{0.45}$ above and below $T_c$.  Over much of the normal state the
material is a weakly-interacting Fermi liquid and the transport is Drude-like.  However,
close to $T_c$ the Drude picture breaks down and the scattering rate adopts a strong
frequency dependence, signalling the increasingly correlated nature of this
material.\cite{qazilbash09,tamai10}
The onset of superconductivity is clearly observed in the optical properties below
$T_c$, and the optical conductivity suggests that in addition to a prominent gap
feature at $\simeq 5.1$~meV, a second gap opens at $\simeq 2.5$~meV.

\begin{figure}[t]
%
%
\centerline{\includegraphics[width=2.7in]{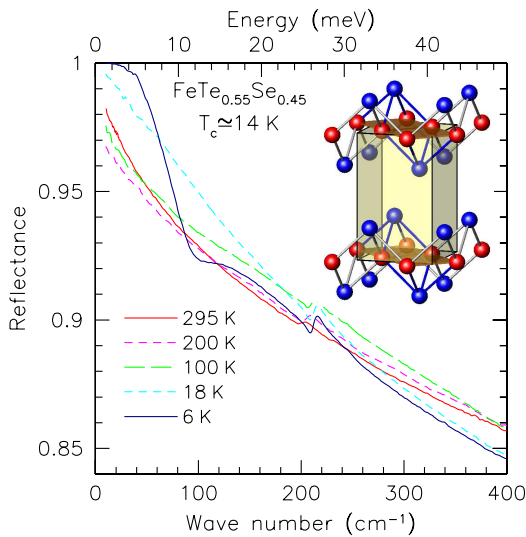}}%
\caption{(Color online)  The reflectance of FeTe$_{0.55}$Se$_{0.45}$ in the far
infrared region for light polarized in the Fe-Te planes at several temperatures
above and below $T_c$.  Inset: The extended unit cell of FeTe in the
tetragonal P4/nmm space group showing the tetrahedrally-coordinated Te above
and below the Fe planes.}
\label{fig:reflec}
\end{figure}

%
%
Single crystals with good cleavage planes (001) were grown by a unidirectional
solidification method with a nominal composition of FeTe$_{0.55}$Se$_{0.45}$ and a
critical temperature determined by magnetic susceptibility of $T_c = 14$~K with
a transition width of $\simeq 1$~K.
The reflectance from the cleaved surface of a mm-sized single crystal
has been measured at a near-normal angle of incidence for several temperatures
above and below $T_c$ over a wide frequency range ($\sim 2$~meV to 4~eV) for
light polarized in the {\em a-b} planes using an {\em in situ} overcoating
technique.\cite{homes93}
The reflectance in the terahertz and far-infrared region (1~THz = 33.4~cm$^{-1}$)
is shown in Fig.~\ref{fig:reflec} (the extended unit cell of FeTe is shown in the
inset).
At room temperature, the reflectance displays the typical metallic form in the
Hagen-Rubens regime $R\propto 1-\sqrt{\omega}$; however, just above $T_c$ the
reflectance develops a striking linear frequency dependence.  Below $T_c$ the formation
of a superconducting condensate and the opening of a gap in the spectrum of excitations
is clearly visible.  The reflectance is a complex quantity consisting
of an amplitude and a phase, $\tilde{r} = \sqrt{R}e^{i\theta}$; because only the
amplitude $R = \tilde{r}\tilde{r}^*$ is measured it is often not intuitively obvious
what changes in the reflectance imply.  Consequently, the complex optical properties
have been determined from a Kramers-Kronig analysis of the reflectance.\cite{dressel-book}

%
%

\begin{figure}[b]
%
%
\centerline{\includegraphics[width=2.7in]{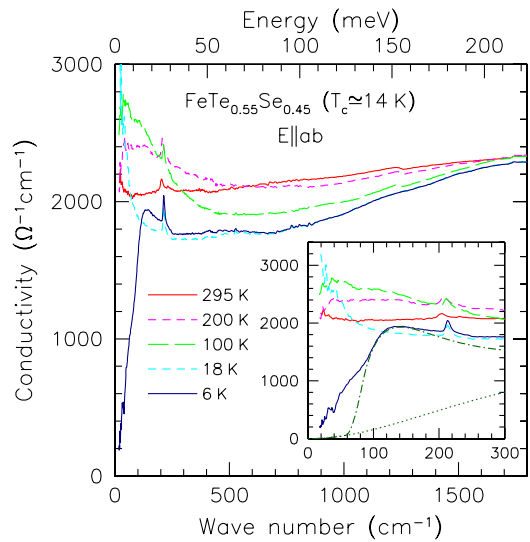}}%
\caption{(Color online) The real part of the in-plane optical conductivity for
FeTe$_{0.55}$Se$_{0.45}$ at several temperatures above and below $T_c$ in the
infrared region.  Inset: The conductivity in the far-infrared region
compared with a model calculation for a single isotropic gap ($\Delta_0 = 4.5$~meV,
$1/\tau = 4\Delta_0$) superimposed on the Lorentzian contribution.}
\label{fig:sigma}
\end{figure}

%
%
The temperature dependence of the real part of the optical conductivity is shown in
Fig.~\ref{fig:sigma} in the infrared region; the far-infrared region
is shown in the inset.  At room temperature, the conductivity is relatively flat
and structureless, except for a sharp feature associated with the infrared-active
$E_u$ mode at 204~cm$^{-1}$ which is due to the in-plane displacements
of the Fe-Te(Se) atoms\cite{xiatl09} (slightly higher than the $E_u$ mode observed
at 187~cm$^{-1}$ in our examination of Fe$_{1.03}$Te).
As the temperature is lowered there is a redistribution of the spectral weight
[defined here as the weight under the conductivity curve over a given interval,
$\int_{0^+}^{\omega_c} \sigma_1(\omega, T)\,d\omega$] from high to low frequency.
This response is not unusual for a metallic system where the scattering rate decreases
with temperature.  The optical conductivity is described by a Drude-Lorentz model
for the dielectric function $\tilde\epsilon=\epsilon_1 + i\epsilon_2$,
\begin{equation}
  \tilde\epsilon(\omega) = \epsilon_\infty - {{\omega_{p,D}^2}\over{\omega^2+i\omega/\tau_D}}
    + \sum_j {{\Omega_j^2}\over{\omega_j^2 - \omega^2 - i\omega\gamma_j}},
\end{equation}
where $\epsilon_\infty$ is the real part of the dielectric function at high
frequency, $\omega_{p,D}^2 = 4\pi ne^2/m^\ast$ and $1/\tau_D$ are the plasma frequency
and scattering rate for the delocalized (Drude) carriers, respectively; $\omega_j$,
$\gamma_j$ and $\Omega_j$ are the position, width, and strength of
the $j$th vibration or excitation.
The complex conductivity is $\tilde\sigma(\omega) = \sigma_1 +i\sigma_2 =
-i\omega [\tilde\epsilon(\omega) - \epsilon_\infty ]/4\pi$.

%
%

The optical conductivity may be reproduced quite well using this approach at
295, 200 and 100~K, with fitted values of $\omega_{p,D} = 7200$~cm$^{-1}$ and
$1/\tau_D = 414, 363$ and 317~cm$^{-1}$, respectively ($\pm 5$\%).  To fit the
midinfrared component, Lorentzian oscillators at the somewhat arbitrary
positions of 650 and 3000~cm$^{-1}$ have been introduced, allowing the
free-carrier component to be fit using a single Drude expression, as
opposed to the two-Drude response that has recently been applied
to some of the pnictide materials.\cite{wu10}  While this approach
works well over most of the normal state, it fails for the optical
conductivity just above $T_c$ at 18~K, where the low-frequency component is not
Drude-like.  To address this problem, we consider the extended-Drude model in
which the scattering rate takes on a frequency dependence.
The experimentally-determined scattering rate is\cite{puchkov96}
\begin{equation}
  {{1}\over{\tau(\omega)}} = {{\omega_p^2}\over{4\pi}} \,
  {\rm Re} \left[ {{1}\over{\tilde\sigma(\omega)}} \right].
\end{equation}
%

%
%
\begin{figure}[t]
%
\centerline{\includegraphics[width=2.7in]{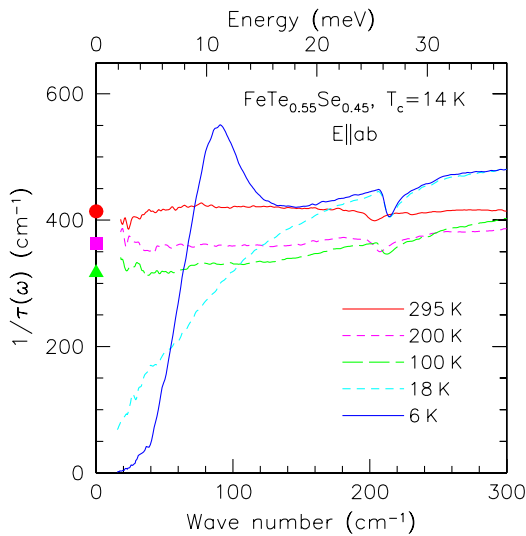}}%
\caption{(Color online) The in-plane frequency-dependent scattering rate
of FeTe$_{0.55}$Se$_{0.45}$ for several temperatures above and below $T_c$
in the far-infrared region.  The values for $1/\tau_D$ are shown at 295 ({\CIRCLE}),
200 ($\blacksquare$) and 100~K ($\blacktriangle$), respectively, where the
scattering rate displays little temperature dependence.  For $T \gtrsim T_c$
(18~K) at low frequency $1/\tau(\omega) \propto \omega$, while for $T < T_c$
large changes in the scattering rate are observed in response to the formation of
superconducting gap(s).
}%
\label{fig:tau}
\end{figure}
%

%
%
In this instance we set $\omega_p = \omega_{p,D}$ and $\epsilon_\infty = 4$
[although the choice of $\epsilon_\infty$ has little effect upon $1/\tau(\omega)$
in the far-infrared region]; the temperature dependence of $1/\tau(\omega)$
is shown in Fig.~\ref{fig:tau} above and below $T_c$.
At 295, 200 and 100~K the scattering rate displays little frequency dependence,
and moreover $1/\tau(\omega\rightarrow 0) \simeq 1/\tau_D$.  This self-consistent
behavior indicates that within this temperature range, the transport may be
described as a weakly-interacting Fermi liquid (Drude model).  However, just
above $T_c$ at 18~K the scattering rate develops a linear frequency
dependence $\lesssim 200$~cm$^{-1}$, suggesting strong electronic
correlations.\cite{tamai10}  This may be due in part to magnetic
correlations\cite{wen09} that arise from the suppression of the magnetic
transition in Fe$_{1+\delta}$Te at $T_N \simeq 70$~K in response to Se
substitution.\cite{sales09}  We note that similar behavior of the scattering
rate is observed in optimally-doped cuprates where the electronic
correlations may have a similar origin.\cite{basov05}
Dramatic changes are also observed in $1/\tau(\omega)$ below $T_c$ where the
scattering rate is suppressed at low frequencies, but increases rapidly and
overshoots the normal-state (18~K) value at about 60~cm$^{-1}$, finally merging
with the normal-state curve at about 200~cm$^{-1}$; this behavior is in rough
agreement with a recently proposed sum rule for the scattering-rate.\cite{basov02}
%
%

%
%
Returning to the optical conductivity in Fig.~\ref{fig:sigma}, below $T_c$ there is
a dramatic suppression of the low-frequency conductivity and a commensurate loss of
spectral weight which is shown in more detail in the inset.  The loss of spectral
weight is associated with the formation of a superconducting condensate, whose strength
may be calculated from the Ferrell-Glover-Tinkham sum rule: $\int_{0^+}^{\omega_c}
\left[ \sigma_1(\omega, T \gtrsim T_c) - \sigma_1(\omega, T\ll T_c) \right] d\omega  =
\omega_{p,S}^2/8$.  Here $\omega_{p,S}^2 = 4\pi n_s e^2/m^*$ is the superconducting
plasma frequency and the cut-off frequency $\omega_c\simeq 150$~cm$^{-1}$ is chosen
so that the integral converges smoothly;
the superfluid density is $\rho_{s0} \equiv \omega_{p,S}^2$.
The sum rule yields $\omega_{p,S} = 3000\pm 200$~cm$^{-1}$, indicating that less
than one-fifth of the free-carriers in the normal state have condensed
($\omega_{p,S}^2/\omega_{p,D}^2 \lesssim 0.18$).  The superfluid density
can also be expressed as an effective penetration depth $\lambda_0 = 5300\pm 300$~\AA ,
which is in good agreement with recent tunnel-diode measurement on
FeTe$_{0.63}$Se$_{0.37}$ (Ref.~\onlinecite{kim10}).
From the estimate $\sigma_{dc}\equiv\sigma_1(\omega\rightarrow 0) =
3500\pm 400$~$\Omega^{-1}{\rm cm}^{-1}$ for $T \gtrsim T_c$, this compound
is observed to fall on the general scaling line\cite{homes04,*homes05} for
a BCS superconductor with the condition that $1/\tau \gtrsim 2\Delta$
(the ``dirty limit''), $\rho_{s0}/8\simeq 8.1\,\sigma_{dc}T_c$.
%
%
%
\begin{figure}[b]
%
\centerline{\includegraphics[width=2.7in]{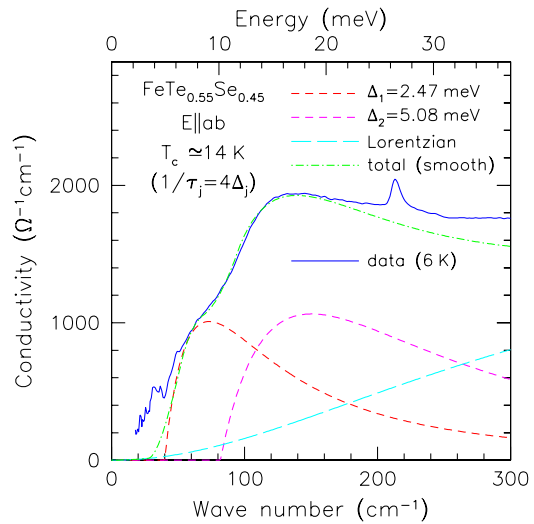}}%
\caption{(Color online) The in-plane optical conductivity of
FeTe$_{0.55}$Se$_{0.45}$ shown at 6~K (solid line).  The calculated
optical conductivity with gaps of $2\Delta_1 \simeq 5$~meV and
$2\Delta_2 \simeq 10.2$~meV for $T \ll T_c$ (short-dashed lines)
is superimposed on the contribution from the bound excitations in
the mid-infrared (long-dashed line); the smoothed linear combination
of the three curves (dot-dash line) is in good agreement with the
measured data below 200~cm$^{-1}$.
}%
\label{fig:detail}
\end{figure}

The detailed optical conductivity below $T_c$ at 6~K is shown in
Fig.~\ref{fig:detail}.  In addition to the strong suppression of the conductivity
below $\sim 120$~cm$^{-1}$, there is also a prominent shoulder at $\sim 60$~cm$^{-1}$.
Below $T_c$ the optical conductivity has been calculated using a
Mattis-Bardeen formalism for the contribution from the gapped
excitations,\cite{zimmerman91,dressel-book} as well as the low-frequency
tail of the bound midinfrared excitations modeled by Lorentzian oscillators.
The Mattis-Bardeen approach assumes that $l\lesssim \xi_0$, where the
mean-free path $l = v_F \tau$ ($v_F$ is the Fermi velocity), and the coherence
length is $\xi_0 = \hbar v_F/\pi\Delta_0$ for an isotropic superconducting gap
$\Delta_0$; this may also be expressed as $1/\tau \gtrsim 2\Delta_0$.
This approach is motivated by the observation that less than one-fifth of the
free carriers collapse into the condensate, a condition which indicates that
these materials are not in the clean limit.
Initially, only a single isotropic gap $\Delta_0 \simeq 4.5$~meV
was considered; however, this failed to accurately reproduce the residual
conductivity observed at low-frequency (inset in Fig.~\ref{fig:sigma}).
To properly model the optical conductivity, two gaps at $\Delta_1 \simeq 2.5$~meV
and $\Delta_2 \simeq 5.1$~meV are used.  For the purposes of this calculation we
have assumed a moderate amount of disorder scattering, $1/\tau_j = 4\Delta_j$.
The optical conductivity for each of the gaps is shown in Fig.~\ref{fig:detail}
for $T=0$; the smoothed linear combination of the gaps and the Lorentzian
tails is in good agreement with the experimental data.
The observation of two gap features is consistent with a number of recent
theoretical works that propose that {\em s}-wave gaps form on each band,
possibly with a sign change between them ($s^\pm$), in this model the
gap on the electron band may be an extended {\em s}-wave with
nodes.\cite{mazin08,chubukov09}  The strong reduction of the conductivity
at low frequency for $T \ll T_c$ suggests the absence of nodes.
It is possible that disorder may lift the nodes, resulting in an nodeless
extended {\em s}-wave gap.\cite{mishra09,carbotte10}  The optical results
provide estimates of the gap amplitudes, but do not distinguish between
$s^\pm$ and extended {\em s}-wave.
%
%
The optical gaps $2\Delta_j \simeq 40$ and 82~cm$^{-1}$ are similar to or
larger than the low-frequency scattering rate observed at 18~K,
$1/\tau(\omega \rightarrow 0) \simeq 40$~cm$^{-1}$. While this
might seem to cast doubt on the validity of the Mattis-Bardeen approach,
we note that the strong frequency dependence of the scattering rate at
this temperature complicates matters. If we consider the value of the
$1/\tau(\omega)$ in the region of the optical gaps $2\Delta_j$ where the
scattering should be important, we find from Fig.~\ref{fig:tau} that the
scattering rate is then larger than the gap amplitude,
$$
 { {1/\tau_j(2\Delta_j)} \over {2\Delta_j} } \approx 3,
$$
which is actually larger than the ratio used in the calculation,
indicating that the Mattis-Bardeen approach is correct.
Finally, we note that while $2\Delta_1/k_{\rm B}T_c \simeq 4$ is close to the
value of 3.5 expected in the BCS weak-coupling limit,\cite{bcs}
$2\Delta_2/k_{\rm B}T_c \simeq 8.4$ is significantly larger.

%
%
To summarize, the optical properties of FeTe$_{0.55}$Se$_{0.45}$ ($T_c = 14$~K)
have been examined for light polarized in the Fe-Te(Se) planes above and below
$T_c$.  Well above $T_c$ the transport may be described by a weakly-interacting
Fermi liquid (Drude model); however, this picture breaks down close to $T_c$
when the scattering rate takes on a strong frequency dependence, similar to
what is observed in the cuprate superconductors.
Below $T_c$, less than one-fifth of the free carriers collapse into the
condensate ($\lambda_0 \simeq 5300$~\AA ), indicating that this material is
in the dirty limit, and indeed this material falls on the general scaling
line predicted for a BCS dirty-limit superconductor.
To successfully model the optical conductivity, two gaps of $\Delta_1 \simeq
2.5$~meV and $\Delta_2 \simeq 5.1$~meV are considered using a Mattis-Bardeen
formalism (with moderate disorder scattering), suggesting either an $s^\pm$ or
an nodeless extended {\em s}-wave gap.

%
%
We would like to acknowledge useful discussions with D. N. Basov, J. P.
Carbotte, A. V. Chubukov and J. M. Tranquada.
This work is supported by the Office of Science, U.S. Department of Energy (DOE)
under Contract No. DE-AC02-98CH10886.

%
%
%
\bibliography{fetese}

\end{document}